\documentclass[10pt,twocolumn,amsmath,amssymb,aps,prb,showpacs,longbibliography,superscriptaddress]{revtex4-1}
\usepackage[latin9]{inputenc}
\usepackage{amsmath}
\usepackage{amssymb}
\usepackage{graphicx}
\usepackage{esint}

\makeatletter

\usepackage{amsfonts}
\usepackage{graphics}

\makeatother

\begin{document}
\title{Finite-momentum bound pairs of two electrons in an altermagnetic metal}
\author{Hui Hu}
\affiliation{Centre for Quantum Technology Theory, Swinburne University of Technology,
Melbourne 3122, Australia}
\author{Zhao Liu}
\affiliation{Centre for Quantum Technology Theory, Swinburne University of Technology,
Melbourne 3122, Australia}
\author{Jia Wang}
\affiliation{Centre for Quantum Technology Theory, Swinburne University of Technology,
Melbourne 3122, Australia}
\author{Xia-Ji Liu}
\affiliation{Centre for Quantum Technology Theory, Swinburne University of Technology,
Melbourne 3122, Australia}
\date{\today}
\begin{abstract}
We solve the two-electron problem on a square lattice with $d$-wave
altermagnetism, considering both on-site and nearest-neighbor attractive
interactions. The altermagnetic spin-splitting in the single-particle
dispersion naturally gives rise to a ground state of two-electron
bound pairs with nonzero center-of-mass momentum. The emergence of
finite-momentum bound states suggests that altermagnetic spin splitting
may favor pairing at nonzero center-of-mass momentum, which could
be relevant for proposed Fulde--Ferrell--Larkin--Ovchinnikov (FFLO)
superconductivity in altermagnetic systems. Additionally, when the
nearest-neighbor attraction is strong, the resulting finite-momentum
bound pairs exhibit a mixture of both spin-singlet and spin-triplet
characteristics, suggesting the possibility of unconventional superconductors,
where spin-singlet and spin-triplet pairings coexist.
\end{abstract}
\maketitle

\section{Introduction}

Exact solutions for few-particle systems often yield valuable insight
into complex quantum many-body problems. A notable example is the
use of the exact two-electron solution of the parent $t$-$J$ model
to understand bound electron pairs in the hole-rich phase of high-$T_{c}$
superconductors \citep{Lin1991,Petukhov1992}, where a low-density
electron superfluid may form if there is an intrinsic tendency toward
pairing. Another important example comes from the exact solutions
of three interacting fermions at the unitary limit (where the $s$-wave
scattering length diverges) \citep{Liu2009,Liu2010a,Liu2010b}, which
provide precise universal thermodynamic behavior at high temperatures
through the quantum virial expansion \citep{Hu2007NatPhys}. In this
work, we present the exact solutions to the two-electron problem in
the presence of $d$-wave altermagnetism \citep{Smejkal2022,Jungwirth2025,Liu2025Review}
and show that these solutions offer a two-body perspective for understanding
the recently proposed altermagnetism-induced Fulde-Ferrell-Larkin-Ovchinnikov
(FFLO) superconducting state and its associated pairing symmetry \citep{Zhang2024,Hu2025PRB}.

Altermagnetism is a newly recognized class of magnetic order that
combines key features of ferromagnetism and antiferromagnetism \citep{Noda2016,Naka2019,Hayami2019,Ahn2019,Hayami2020,Smejkal2020,Mazin2023,Amin2024,Wang2026}.
Although the net magnetization in altermagnetic materials is zero
due to their collinearly alternating spins on different sublattices,
specific crystal symmetries produce a pronounced, momentum-dependent
spin splitting in the electronic band structure even without an external
magnetic field. As a result, altermagnets can support spin-polarized
electronic bands similar to those found in ferromagnets, enabling
exotic quantum phases such as FFLO superconductivity \citep{Fulde1964,Larkin1964,Casalbuoni2004,Hu2006,Hu2007}.
This altermagnetism-driven FFLO state has been the subject of detailed
studies in various contexts, including systems with purely $s$-wave
on-site ($U$) and with higher-order partial-wave nearest-neighbor
($V$) attractive interactions \citep{Zhang2024,Hu2025PRB,Sumita2023,Chakraborty2024,Fukaya2025,Hong2025,Iorsh2025,Sim2024,Sumita2025,Hu2025AB,Liu2026PRB,Liu2026AB,Jasiewicz2026},
without or with spin-orbit coupling \citep{deCarvalho2024,Mukasa2025}.
In contrast to conventional FFLO mechanisms that rely on external
Zeeman fields, spin-orbit coupling, or other sources of Fermi-surface
mismatch, the altermagnetism-induced FFLO state originates from the
intrinsic momentum-dependent spin splitting characteristic of altermagnetic
order. The finite pairing momentum emerges directly from the anisotropic
spin-split electronic structure, without requiring magnetic-field-induced
polarization. Consequently, altermagnetism provides a unique and fundamentally
distinct platform for realizing FFLO superconductivity.

In the present study, we take a few-body approach to altermagnetism-induced
FFLO superconductivity by exactly solving the two-electron bound-pair
problem using an extended Hubbard $U$-$V$ model in two dimensions.
At strong altermagnetic coupling, the two-electron continuum is significantly
reshaped, developing its lowest energy at a nonzero center-of-mass
momentum. As a result, the attractive interactions stabilize a ground
state of two-electron bound pairs with finite momentum, providing
the two-electron analog of many-electron FFLO pairing observed in
altermagnetic systems. Moreover, we find that altermagnetism strongly
alters the pairing symmetry of these bound pairs: they can no longer
be cleanly classified as purely spin-singlet or spin-triplet, but
instead generally exhibit mixed characteristics. In the many-electron
case, this suggests the possible emergence of superconducting states
featuring coherent superpositions of spin-singlet and spin-triplet
pairings.

The remainder of the paper is organized as follows. In the next section
(Sec. II), we briefly introduce the model Hamiltonian. In Sec. III,
we present the two-electron ansatz and outline the method for solving
the bound states below the two-electron continuum. In Sec. IV, focusing
on the case of $d_{xy}$-wave altermagnetism, we discuss the bound-pair
energy spectrum and the pairing symmetry of the associated wave functions.
In Sec. V, we summarize our conclusions and discuss potential future
directions. Moreover, Appendix A examines the general behavior of
the whole two-electron spectrum, and Appendix B provides a concise
summary of results for $d_{x^{2}-y^{2}}$-wave altermagnetism.

\section{Model Hamiltonian}

We consider two electrons moving on a square lattice with the nearest-neighbor
hopping strength $t$. In the presence of the $d$-wave altermagnetism,
the model Hamiltonian can be written as, $\mathcal{H}=\mathcal{H}_{0}+\mathcal{H}_{\textrm{int}}$,
where for convenience the non-interacting kinetic part $\mathcal{H}_{0}$
in momentum space takes the simplest one-band form,
\begin{equation}
\mathcal{H}_{0}=\sum_{\mathbf{k}}\left[\left(\varepsilon_{\mathbf{k}}+J_{\mathbf{k}}\right)c_{\mathbf{k}\uparrow}^{\dagger}c_{\mathbf{k}\uparrow}+\left(\varepsilon_{\mathbf{k}}-J_{\mathbf{k}}\right)c_{\mathbf{k}\downarrow}^{\dagger}c_{\mathbf{k}\downarrow}\right],
\end{equation}
with the single-particle dispersion relation $\varepsilon_{\mathbf{k}}\equiv-2t(\cos k_{x}+\cos k_{y})$
and the the momentum-dependent spin-splitting $J_{\mathbf{k}}$ arising
from altermagnetism. Here and in the following, $c_{\mathbf{k}\sigma}^{\dagger}$
and $c_{i\sigma}^{\dagger}$ represent the creation operators at the
momentum $\mathbf{k}$ and at the site $i$, respectively, for either
spin-up ($\sigma=\uparrow$) or spin-down ($\sigma=\downarrow$) electrons.
We will primarily focus on the case of the $d_{xy}$-wave altermagnetism,
where $J_{\mathbf{k}}$ is given by,
\begin{equation}
J_{\mathbf{k}}=\lambda\sin\left(k_{x}\right)\sin\left(k_{y}\right).
\end{equation}
As a minimal model, this altermagnetic order can be realized through
spin-dependent next-nearest-neighbor hopping processes on a square
lattice, rather than through nearest-neighbor hoppings along the $x$-
and $y$-directions. The hopping amplitudes are assigned opposite
signs on the diagonal and anti-diagonal bonds, giving rise to a characteristic
$d_{xy}$-wave pattern \citep{Hong2025}. Owing to such an alternating
sign structure, the altermagnetic hopping term is not invariant under
the four-fold lattice rotation $C_{4}$, which transforms $k_{x}\rightarrow-k_{y}$
and $k_{y}\rightarrow k_{x}$, as the diagonal and anti-diagonal bonds
are interchanged. However, the sign change induced by the $C_{4}$
rotation is exactly compensated by the spin-dependent nature of the
hopping term under a spin-flip operation. Consequently, although neither
$C_{4}$ nor time-reversal symmetry $\mathcal{T}$ is individually
preserved, the Hamiltonian remains invariant under their combined
operation $C_{4}\mathcal{T}$. This combined spin-lattice symmetry
is a defining feature of the altermagnetic state and underlies the
emergence of its characteristic momentum-dependent spin splitting.
In Appendix B, we will also consider the alternative case of the $d_{x^{2}-y^{2}}$-wave
altermagnetism, where
\begin{equation}
J_{\mathbf{k}}=\frac{\lambda}{2}\left[\cos k_{x}-\cos k_{y}\right].
\end{equation}
To simplify the notation, throughout the paper we denote $\varepsilon_{\mathbf{k}\uparrow}=\varepsilon_{\mathbf{k}}+J_{\mathbf{k}}$
and $\varepsilon_{\mathbf{k}\downarrow}=\varepsilon_{\mathbf{k}}-J_{\mathbf{k}}$.

\subsection{Interaction Hamiltonian}

For the interaction Hamiltonian, we consider the short-range interaction
potential on a $L\times L$ square lattice and use the extended Hubbard
$U$-$V$ model \citep{Kornilovitch2004,Kornilovitch2014,Kornilovitch2024},
with the on-site ($U$) and the nearest-neighbor interaction strengths
($V$):

\begin{equation}
\mathcal{H}_{\textrm{int}}=\frac{U}{2}\sum_{i}n_{i}\left(n_{i}-1\right)+\frac{V}{2}\sum_{i\delta}n_{i}n_{i+\delta},
\end{equation}
where $\delta=-\hat{x},+\hat{x},-\hat{y},+\hat{y}$ denotes the four
nearest-neighbor sites on the square lattice, and $n_{i}=n_{i\uparrow}+n_{i\downarrow}=\sum_{\sigma}c_{i\sigma}^{\dagger}c_{i\sigma}$
is the number of electrons on the $i$-th lattice site. It is readily
seen that the on-site interaction occurs only between spin-up and
spin-down electrons, since $n_{i\sigma}^{2}=n_{i\sigma}$. In contrast,
the nearest-neighbor interaction can happen between electrons with
same spins. In realistic electronic systems, both the on-site and
nearest-neighbor interactions arise from the Coulomb repulsion. In
the low-energy limit, however, the strong on-site repulsion gives
rise to substantial spin fluctuations, resulting in a relatively strong
effective nearest-neighbor attraction \citep{Hirsch1985,Scalapino1986}.
Unless otherwise specified, we focus on the parameter regime where
the effective interaction strengths satisfy $\left|V\right|>U$.

In momentum space, after the Fourier transformation we may rewrite
the interaction Hamiltonian in the form \citep{Hong2025,Zhu2023},
\begin{eqnarray}
\mathcal{H}_{\textrm{int}} & = & \frac{1}{\mathcal{S}}\sum_{\mathbf{k},\mathbf{k}',\mathbf{q};\sigma}V_{\mathbf{k},\mathbf{k}'}^{\left(\sigma\right)}c_{\mathbf{k}+\frac{\mathbf{q}}{2}\sigma}^{\dagger}c_{\mathbf{-k}+\frac{\mathbf{q}}{2}\sigma}^{\dagger}c_{\mathbf{-k'}+\frac{\mathbf{q}}{2}\sigma}c_{\mathbf{k}'+\frac{\mathbf{q}}{2}\sigma}+\nonumber \\
 &  & \frac{1}{\mathcal{S}}\sum_{\mathbf{k},\mathbf{k}',\mathbf{q}}V_{\mathbf{k},\mathbf{k}'}^{\left(\uparrow\downarrow\right)}c_{\mathbf{k}+\frac{\mathbf{q}}{2}\uparrow}^{\dagger}c_{\mathbf{-k}+\frac{\mathbf{q}}{2}\downarrow}^{\dagger}c_{\mathbf{-k'}+\frac{\mathbf{q}}{2}\downarrow}c_{\mathbf{k}'+\frac{\mathbf{q}}{2}\uparrow},
\end{eqnarray}
where $\mathcal{S}=L^{2}$ is the total number of the lattice sites.
The two terms describe the interactions between electrons with same
spin and opposite spin, respectively, with the interaction strengths
\citep{Hong2025,Zhu2023}:
\begin{eqnarray}
V_{\mathbf{k},\mathbf{k}'}^{\left(\sigma\right)} & = & V\left[\cos\left(k_{x}-k'_{x}\right)+\cos\left(k_{y}-k'_{y}\right)\right],\\
V_{\mathbf{k},\mathbf{k}'}^{\left(\uparrow\downarrow\right)} & = & U+2V\left[\cos\left(k_{x}-k'_{x}\right)+\cos\left(k_{y}-k'_{y}\right)\right].\label{eq: Vkkp}
\end{eqnarray}
The on-site interaction is purely $s$-wave and does not depend on
momentum, whereas the nearest-neighbor interaction exhibits significant
momentum dependence. This nearest-neighbor term can be decomposed
into distinct channels (i.e., the extended $s$-wave, the two $p$-wave,
and the $d$-wave channels), each expressible in a \emph{separable}
form. In particular, we can rewrite \citep{Hong2025}
\begin{equation}
V_{\mathbf{k},\mathbf{k}'}^{\left(\uparrow\downarrow\right)}=\sum_{\eta=\textrm{s,es,p+ip,p-ip,d}}V_{\eta}f_{\eta}\left(\mathbf{k}\right)f_{\eta}^{*}\left(\mathbf{k}'\right),
\end{equation}
where $V_{\textrm{s}}=U$, $V_{\textrm{es}}=V_{\textrm{p+ip}}=V_{\textrm{p-ip}}=V_{\textrm{d}}=V$,
and the form factors of different channels are given by,
\begin{eqnarray}
f_{\textrm{s}}\left(\mathbf{k}\right) & = & 1,\\
f_{\textrm{es}}\left(\mathbf{k}\right) & = & \cos k_{x}+\cos k_{y},\\
f_{\textrm{p+ip}}\left(\mathbf{k}\right) & = & \sin k_{x}+i\sin k_{y},\\
f_{\textrm{p-ip}}\left(\mathbf{k}\right) & = & \sin k_{x}-i\sin k_{y},\\
f_{\textrm{d}}\left(\mathbf{k}\right) & = & \cos k_{x}-\cos k_{y}.
\end{eqnarray}
We can clearly identify the parity-even channel ($\eta=\textrm{s},\textrm{es},\textrm{d}$
for the spin-singlet pairing) and the parity-odd channel ($\eta=\textrm{p+ip},\textrm{p-ip}$
for the spin-triplet pairing). 

In numerical calculations, to avoid the use of the complex number,
it is more convenient to adopt the alternative form factors for the
two $p$-wave channels ($\eta=p+,p-$):
\begin{eqnarray}
f_{\textrm{\ensuremath{p+}}}\left(\mathbf{k}\right) & = & \sin k_{x}+\sin k_{y},\\
f_{p-}\left(\mathbf{k}\right) & = & \sin k_{x}-\sin k_{y}.
\end{eqnarray}
This choice is particularly appropriate for $d_{xy}$-wave altermagnetism,
where the band splitting reaches its maximum along the diagonal and
anti-diagonal directions ($k_{x}=\pm k_{y}$). In this case, the $p+$
and $p-$ basis functions used for $p$-wave channels are naturally
aligns with the symmetry directions of the band splitting. By contrast,
for $d_{x^{2}-y^{2}}$-wave altermagnetism, where the band splitting
is largest along the $k_{x}$ and $k_{y}$ axes, it is more physically
intuitive to employ the conventional unrotated $p$-wave form factors,
$f_{p_{x}}(\mathbf{k})=\sqrt{2}\sin k_{x}$ and $f_{p_{y}}(\mathbf{k})=\sqrt{2}\sin k_{y}$.
Although different choices of $p$-wave basis functions yield the
same energy spectrum, they can provide a more transparent characterization
of the corresponding eigenstates.

It should be noted that, one can easily extend the interaction range
and incorporate additional terms like next-nearest-neighbor interactions
to more accurately represent a non-local interaction potential. This
extension generates higher-order partial-wave components in both $V_{\mathbf{k},\mathbf{k}'}^{(\sigma)}$
and $V_{\mathbf{k},\mathbf{k}'}^{(\uparrow\downarrow)}$.

\section{Theoretical framework}

For concreteness, we focus on the case of two electrons with \emph{opposite}
spins. The same-spin scenario is less relevant here, since altermagnetism
primarily alters the single-particle dispersion relation in a spin-independent
way rather than affecting same-spin interactions. Therefore, only
Eq. (\ref{eq: Vkkp}) contributes here.

\subsection{Exact diagonalization of a finite lattice system}

For a small lattice size $L$, it is convenient to diagonalize the
Hamiltonian by employing the following two-electron ansatz at a given
center-of-mass momentum $\mathbf{Q}$:
\begin{equation}
\left|\mathbf{p}\right\rangle \equiv c_{\mathbf{p}+\frac{\mathbf{Q}}{2}\uparrow}^{\dagger}c_{\mathbf{-p}+\frac{\mathbf{Q}}{2}\downarrow}^{\dagger}\left|\textrm{vac}\right\rangle .
\end{equation}
It is straightforward to derive that,
\begin{eqnarray}
\mathcal{H}_{0}\left|\mathbf{p}\right\rangle  & = & \left(\varepsilon_{\mathbf{p}+\frac{\mathbf{Q}}{2}\uparrow}+\varepsilon_{-\mathbf{p}+\frac{\mathbf{Q}}{2}\downarrow}\right)\left|\mathbf{p}\right\rangle ,\label{eq:h0_p}\\
\mathcal{H}_{\textrm{int}}\left|\mathbf{p}\right\rangle  & = & \frac{1}{\mathcal{S}}\sum_{\mathbf{k}}V_{\mathbf{k},\mathbf{p}}^{\left(\uparrow\downarrow\right)}\left|\mathbf{k}\right\rangle .\label{eq:hint_p}
\end{eqnarray}
Therefore, with the complete set of the two-electron ansatz, we obtain
the matrix elements of the Hamiltonian,
\begin{equation}
\left\langle \mathbf{k}\right|\mathcal{H}\left|\mathbf{p}\right\rangle =\delta_{\mathbf{k}\mathbf{p}}\left(\varepsilon_{\mathbf{p}+\frac{\mathbf{Q}}{2}\uparrow}+\varepsilon_{-\mathbf{p}+\frac{\mathbf{Q}}{2}\downarrow}\right)+\frac{1}{\mathcal{S}}V_{\mathbf{k},\mathbf{p}}^{\left(\uparrow\downarrow\right)}.\label{eq:hmatrix}
\end{equation}
By expressing the momentum within the first Brillouin zone, $\mathbf{k}=(k_{x},k_{y})=(n_{x},n_{y})2\pi/L$,
where the integers $n_{x}$ and $n_{y}$ range from $-L/2+1$ to $L/2$
for an even lattice length $L$ and the lattice constant is set to
unity, we can easily diagonalize the Hamiltonian for lattice size
$L$ up to several hundred.

\subsection{Bound states of an infinite lattice system}

To treat the case of an infinite lattice, we can formulate the two-electron
wave-function,
\begin{equation}
\left|\Phi\right\rangle =\sum_{\mathbf{p}}\Phi_{\mathbf{p}}\left|\mathbf{p}\right\rangle =\sum_{\mathbf{p}}\Phi_{\mathbf{p}}c_{\mathbf{p}+\frac{\mathbf{Q}}{2}\uparrow}^{\dagger}c_{\mathbf{-p}+\frac{\mathbf{Q}}{2}\downarrow}^{\dagger}\left|\textrm{vac}\right\rangle .
\end{equation}
 By using Eq. (\ref{eq:h0_p}) and Eq. (\ref{eq:hint_p}), one obtains,
\begin{eqnarray}
\mathcal{H}_{0}\left|\Phi\right\rangle  & = & \sum_{\mathbf{p}}\left(\varepsilon_{\mathbf{p}+\frac{\mathbf{Q}}{2}\uparrow}+\varepsilon_{-\mathbf{p}+\frac{\mathbf{Q}}{2}\downarrow}\right)\Phi_{\mathbf{p}}\left|\mathbf{p}\right\rangle ,\\
\mathcal{H}_{\textrm{int}}\left|\Phi\right\rangle  & = & \frac{1}{\mathcal{S}}\sum_{\mathbf{p}}\sum_{\mathbf{k}}V_{\mathbf{p},\mathbf{k}}^{\left(\uparrow\downarrow\right)}\Phi_{\mathbf{k}}\left|\mathbf{p}\right\rangle .
\end{eqnarray}
By projecting onto the two-electron ansatz, we rewrite the two-electron
Schrödinger equation $(\mathcal{H}_{0}+\mathcal{H}_{\textrm{int}})\left|\Phi\right\rangle =E\left|\Phi\right\rangle $
into the standard form,
\begin{equation}
\left(\varepsilon_{\mathbf{p}+\frac{\mathbf{Q}}{2}\uparrow}+\varepsilon_{-\mathbf{p}+\frac{\mathbf{Q}}{2}\downarrow}\right)\Phi_{\mathbf{p}}+\frac{1}{\mathcal{S}}\sum_{\mathbf{k}}V_{\mathbf{p},\mathbf{k}}^{\left(\uparrow\downarrow\right)}\Phi_{\mathbf{k}}=E\Phi_{\mathbf{p}}.
\end{equation}
To solve this Schrödinger equation, it is useful to insert the separable
form of the interaction potential,
\begin{equation}
\frac{1}{\mathcal{S}}\sum_{\mathbf{k}\eta}V_{\eta}f_{\eta}\left(\mathbf{p}\right)f_{\eta}^{*}\left(\mathbf{k}\right)\Phi_{\mathbf{k}}=\left(E-\varepsilon_{\mathbf{p}+\frac{\mathbf{Q}}{2}\uparrow}-\varepsilon_{-\mathbf{p}+\frac{\mathbf{Q}}{2}\downarrow}\right)\Phi_{\mathbf{p}}.
\end{equation}
At this point, let us introduce the variables
\begin{equation}
M_{\eta}\equiv\frac{1}{\mathcal{S}}\sum_{\mathbf{k}}f_{\eta}^{*}\left(\mathbf{k}\right)\Phi_{\mathbf{k}},
\end{equation}
and solve for the wave-function, 
\begin{equation}
\Phi_{\mathbf{p}}=\frac{\sum_{\eta}V_{\eta}f_{\eta}\left(\mathbf{p}\right)M_{\eta}}{E-\varepsilon_{\mathbf{p}+\frac{\mathbf{Q}}{2}\uparrow}-\varepsilon_{-\mathbf{p}+\frac{\mathbf{Q}}{2}\downarrow}}.\label{eq:wfs}
\end{equation}
The self-consistency leads to a set of coupled equations for the variables
$M_{\eta}$:
\begin{equation}
M_{\eta}=\frac{1}{\mathcal{S}}\sum_{\mathbf{k}}f_{\eta}^{*}\left(\mathbf{k}\right)\frac{\sum_{\eta'}V_{\eta'}f_{\eta'}\left(\mathbf{k}\right)M_{\eta'}}{E-\varepsilon_{\mathbf{k}+\frac{\mathbf{Q}}{2}\uparrow}-\varepsilon_{-\mathbf{k}+\frac{\mathbf{Q}}{2}\downarrow}}.
\end{equation}
Thus, it is useful to define,
\begin{equation}
L_{\eta\eta'}\left(E\right)\equiv\frac{1}{\mathcal{S}}\sum_{\mathbf{k}}\frac{f_{\eta}^{*}\left(\mathbf{k}\right)f_{\eta'}\left(\mathbf{k}\right)}{E-\left(\varepsilon_{\mathbf{k}+\frac{\mathbf{Q}}{2}\uparrow}+\varepsilon_{-\mathbf{k}+\frac{\mathbf{Q}}{2}\downarrow}\right)},
\end{equation}
and rewrite the previous coupled equations into the form,
\begin{equation}
\frac{\left(V_{\eta}M_{\eta}\right)}{V_{\eta}}=\sum_{\eta'}\left[L_{\eta\eta'}\left(E\right)\right]\left(V_{\eta'}M_{\eta'}\right).
\end{equation}
A non-trivial solution for the variables $V_{\eta}M_{\eta}$ requires
the zero determinant \citep{Kornilovitch2004}, 
\begin{equation}
\det\left[L_{\eta\eta'}\left(E\right)-\frac{\delta_{\eta\eta'}}{V_{\eta}}\right]=0,\label{eq:det}
\end{equation}
for a 5 by 5 matrix $\mathbf{A}$, where the matrix elements $A_{\eta\eta'}=L_{\eta\eta'}\left(E\right)-V_{\eta}^{-1}\delta_{\eta\eta'}$.
This secular equation determines all the two-electron energy levels
$E$, including the bound states below the threshold of the two-particle
continuum (for a given center-of-mass momentum $\mathbf{Q}$), i.e.,
\begin{equation}
E<E_{2p}^{(0)}\left(\mathbf{Q}\right)=\min_{\mathbf{k}}\{\varepsilon_{\mathbf{k}+\frac{\mathbf{Q}}{2}\uparrow}+\varepsilon_{-\mathbf{k}+\frac{\mathbf{Q}}{2}\downarrow}\}.
\end{equation}
The eigenvectors of the matrix $\mathbf{A}$, upon substitution in
Eq. (\ref{eq:wfs}), give rise to the wave-function $\Phi_{\mathbf{p}}$
up to a normalization constant.

\subsection{Classification into the spin-singlet and spin-triplet states}

In the absence of altermagnetism, the two-electron dispersion relation
$E_{2p}(\mathbf{k},\mathbf{Q})=\varepsilon_{\mathbf{k}+\mathbf{Q}/2\uparrow}+\varepsilon_{-\mathbf{k}+\mathbf{Q}/2\downarrow}$
can be expressed as follows \citep{Kornilovitch2004},
\begin{equation}
E_{2p}\left(\mathbf{k},\mathbf{Q}\right)=-4t\left(\cos\frac{Q_{x}}{2}\cos k_{x}+\cos\frac{Q_{y}}{2}\cos k_{y}\right),
\end{equation}
which is an even function in momentum $\mathbf{k}$. Recall that the
form factors of the $s$-wave, extend $s$-wave and $d$-wave channels
are even function of $\mathbf{k}$, whereas the two $p$-wave form
factors are odd function. From this it follows that $L_{\eta\eta'}=0$,
whenever the indices $\eta$ and $\eta'$ belong to different parity
classes. As a consequence, the two-electron wave-function have a well-defined
parity. In other words, the two-electron states are purely spin-singlet
or purely spin-triplet, with no admixture of singlet and triplet components.

The situation changes dramatically in the presence of altermagnetism
($\lambda\neq0$). In this case, the two-electron dispersion relation
$E_{2p}(\mathbf{k},\mathbf{Q})$ is expressed as,\begin{widetext}
\begin{equation}
E_{2p}\left(\mathbf{k},\mathbf{Q}\right)=-4t\left(\cos\frac{Q_{x}}{2}\cos k_{x}+\cos\frac{Q_{y}}{2}\cos k_{y}\right)+2\lambda\left(\cos\frac{Q_{x}}{2}\sin\frac{Q_{y}}{2}\sin k_{x}\cos k_{y}+\sin\frac{Q_{x}}{2}\cos\frac{Q_{y}}{2}\cos k_{x}\sin k_{y}\right).\label{eq:E2pLAM}
\end{equation}
\end{widetext}containing both even and odd components of $\mathbf{k}$.
Thus, in general $L_{\eta\eta'}$ need not vanish. As a result, all
the two-electron states exhibit pairing in both the spin-singlet and
spin-triplet channels. This mixed pairing character can be easily
identified through the quantities $M_{\eta}$. For numerical analysis,
it is often useful to introduce the normalized variables, 
\begin{equation}
\tilde{M}_{\eta}=\frac{\left|M_{\eta}\right|}{\sqrt{\sum_{\eta}M_{\eta}^{2}}},
\end{equation}
which serve to characterize the dominant pairing channels of the bound
pairs.

\section{Results and discussions}

The numerical solution of the two-electron bound states using Eq.
(\ref{eq:det}) is straightforward, since the integral $L_{\eta\eta'}(E)$
is well-defined for energies below the two-electron scattering continuum,
i.e., $E<E_{2p}^{(0)}(\mathbf{Q})$. We have also independently examined
the bound states on finite lattices with size $L$ up to 100, and,
as expected, the dependence of the bound-pair energy on lattice size
is exponentially weak. In what follows, we treat separately the two
scenarios in which either the on-site attraction ($U<0$) or the nearest-neighbor
attraction ($V<0$) plays the dominant role. We concentrate on $d_{xy}$-wave
altermagnetism and, in Appendix B, provide a brief discussion of the
results for $d_{x^{2}-y^{2}}$-wave altermagnetism.

\begin{figure}
\begin{centering}
\includegraphics[width=0.5\textwidth]{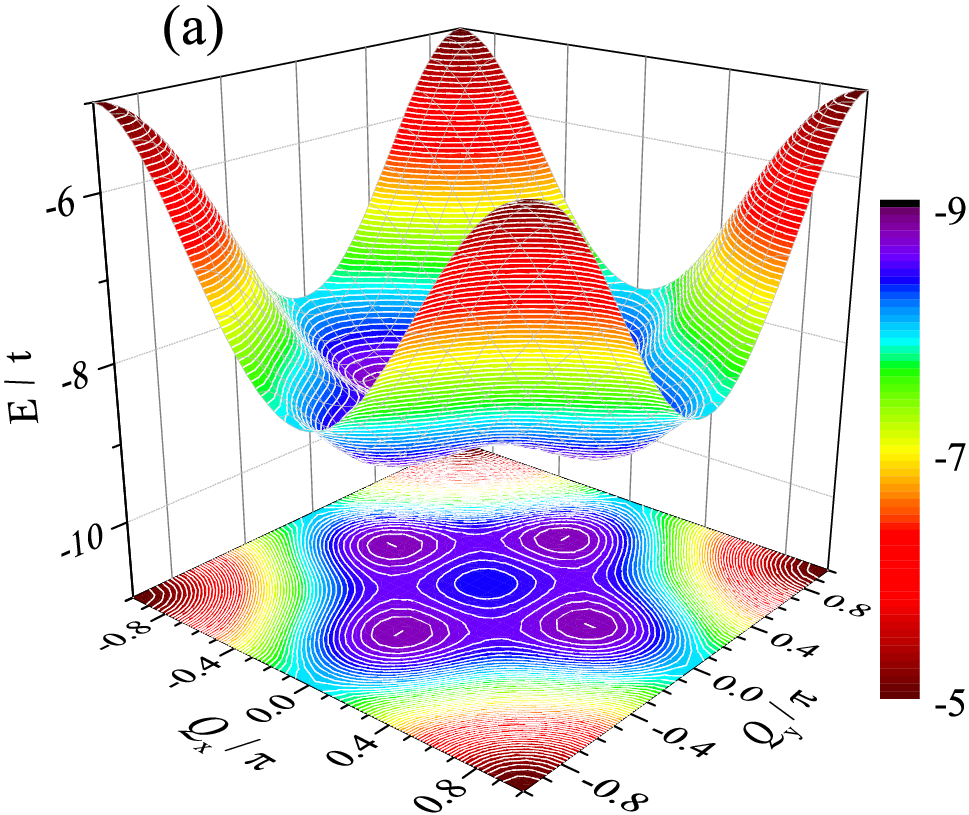}
\par\end{centering}
\begin{centering}
\includegraphics[width=0.5\textwidth]{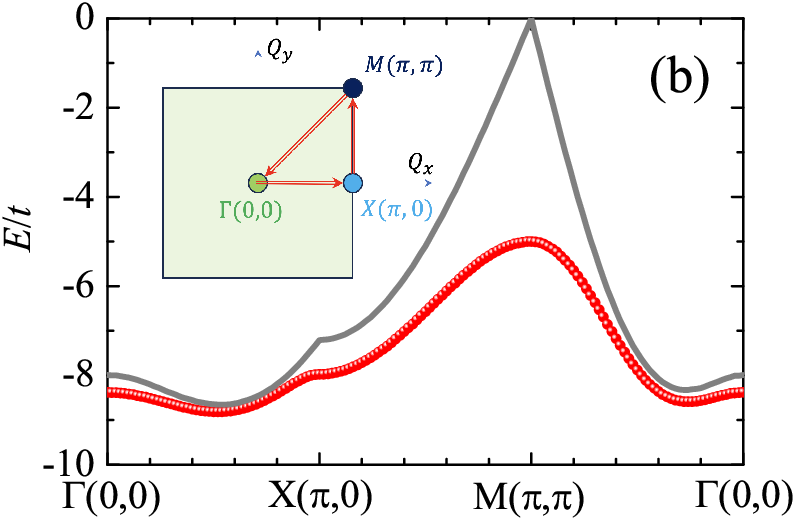}
\par\end{centering}
\caption{\label{fig1} The energy of the bound pairs as a function of the center-of-mass
momentum $\mathbf{Q}=(Q_{x},Q_{y})$, at the on-site attractive interaction
$U=-5t$ with the $d_{xy}$-wave altermagnetic coupling $\lambda=3t$:
(a) a 2D contour plot in the $Q_{x}$-$Q_{y}$ plane, and (b) a 1D
plot along the high-symmetry lines in the first Brillouin zone, as
indicated in the inset. The grey line shows the lower bound of the
two-electron continuum $E_{2p}^{(0)}(\mathbf{Q})$, whose dispersion
features local minima at nonzero momentum due to the presence of altermagnetism.
The bound-state energy closely tracks this continuum edge and therefore
exhibits corresponding local minima at finite momentum in its dispersion.
Here, we set the nearest-neighbor interaction $V=0$. }
\end{figure}

\subsection{The case with only attractive on-site interaction $U<0$}

In Fig. \ref{fig1}, we report the bound-pair energy $E$ as a function
of the center-of-mass momentum $\mathbf{Q}=(Q_{x},Q_{y})$ for an
on-site attractive interaction $U=-5t$. In the presence of a $d_{xy}$-wave
altermagnetic coupling $\lambda=3t$, the two-dimensional (2D) contour
plot in Fig. \ref{fig1}(a) clearly indicates that the ground-state
of the bound pairs occurs at finite momentum along either the $Q_{x}$
or $Q_{y}$ axis. Furthermore, the one-dimensional (1D) cut along
the high-symmetry $\Gamma$-$X$-$M$-$\Gamma$ path in Fig. \ref{fig1}(b)
shows that the bound-pair energy closely tracks the lower edge of
the two-electron continuum $E_{2p}^{(0)}(\mathbf{Q})$ near the $\Gamma$
point. As this continuum threshold also reaches a minimum at nonzero
momentum, the emergence of finite-momentum bound pairs appears to
directly reflect the non-trivial two-electron dispersion induced by
altermagnetism.

\begin{figure}
\begin{centering}
\includegraphics[width=0.5\textwidth]{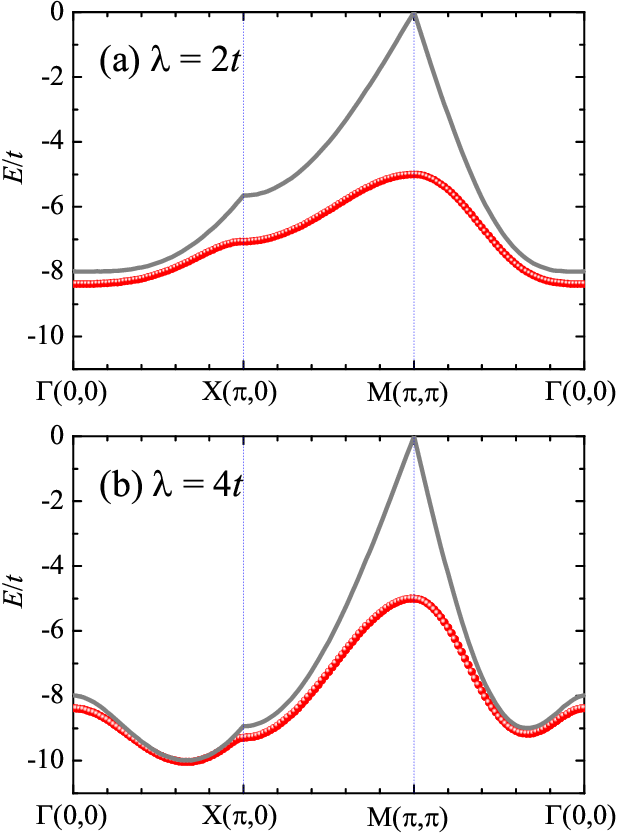}
\par\end{centering}
\caption{\label{fig2} The same as in Fig. \ref{fig1}(b), but with different
$d_{xy}$-wave altermagnetic coupling strengths: $\lambda=2t$ (a)
and $\lambda=4t$ (b). In the former case, corresponding to weak altermagnetic
coupling, the bound-state energy attains its minimum at zero momentum.
In contrast, for sufficiently strong altermagnetic coupling, the bound-state
dispersion develops local minima at finite momentum.}
\end{figure}

To verify this expectation, Fig. \ref{fig2} presents the bound-pair
energy for two additional values of the altermagnetic coupling. For
a weaker coupling ($\lambda=2t$ in Fig. \ref{fig2}(a)), where the
two-electron continuum exhibits its minimum at the $\Gamma$ point,
we indeed find that the ground-state bound pair occurs at zero center-of-mass
momentum. For a stronger coupling ($\lambda=4t$ in Fig. \ref{fig2}(b)),
the ground-state bound pair shares exactly the same finite center-of-mass
momentum at which the two-electron continuum reaches its minimum.

As the minimum of the two-electron continuum plays a crucial role
in determining the emergence of finite-momentum bound states, it is
instructive to analyze its behavior analytically. To this end, we
focus on the high-symmetry $\Gamma$-$M$ line, where $Q_{x}=Q_{y}=Q$.
Introducing the parametrization $k_{x}=-\theta/2+\varphi$ and $k_{y}=-\theta/2-\varphi$,
the two-electron continuum can be expressed as,

\begin{equation}
E_{2p}\left(\mathbf{k},\mathbf{Q}\right)=-8t\cos\frac{Q}{2}\cos\frac{\theta}{2}\cos\varphi-\lambda\sin Q\sin\theta.
\end{equation}
It is straightforward to show that the lower edge of the continuum,
$E_{2p}^{(0)}(\mathbf{Q})$, is attained at $\varphi=0$ and $\theta=Q$,
yielding

\begin{equation}
E_{2p}^{(0)}\left(\mathbf{Q}\right)=-8t\cos^{2}\frac{Q}{2}-\lambda\sin^{2}Q.
\end{equation}
The behavior of this function depends sensitively on the strength
of the altermagnetic coupling $\lambda$. For $\lambda\le2t$, the
continuum minimum remains at $Q=0$, indicating that the lowest-energy
two-particle states carry zero center-of-mass momentum. In contrast,
when $\lambda>2t$, the minimum shifts continuously to a finite momentum
$Q=\arccos(2t/\lambda)$, signaling a tendency toward finite-momentum
pairing. As a representative example, for $\lambda=4t$ one obtains
$Q=\arccos(1/2)=\pi/3$, in excellent agreement with the numerical
results shown in Fig. \ref{fig2}(b). This analysis demonstrates that
the altermagnetic coupling drives a transition of the continuum minimum
from zero to finite momentum at the critical value $\lambda_{c}=2t$,
thereby providing the microscopic origin of the finite-momentum bound
states.

The binding energy $E_{b}$ of the bound state, defined as $E_{b}=E_{2p}^{(0)}(\mathbf{Q})-E$,
generally decreases as the altermagnetic coupling increases. This
behavior is anticipated, since the altermagnetism acts effectively
as a momentum-dependent magnetic field, which ultimately suppresses
pairing and can break bound pairs at sufficiently large altermagnetic
coupling strength. The only exceptions to the overall decrease in
binding energy are at the two high-symmetry points, $\Gamma$ and
$M$, where the binding energy remains unchanged. At these points,
$J_{\mathbf{k}+\mathbf{Q}/2}=J_{-\mathbf{k}+\mathbf{Q}/2}$, so the
effect of altermagnetism cancels exactly. We also note that the bound-pair
energy at the $M$ point equals the on-site interaction $U=-5t$.
This corresponds to the so-called $\eta$-pairing state \citep{Yang1989},
in which the two electrons form a pair on the same lattice site and
are distributed in a staggered pattern throughout the whole lattice.

The existence of a finite-momentum bound-pair ground state points
to a possible propensity for finite-momentum FFLO pairing in the corresponding
many-fermion system, a phenomenon that has been widely explored in
recent theoretical studies, with $s$-wave interaction potential.
In particular, in the case of $d_{xy}$-wave altermagnetism, the FFLO
pairing emerges when the center-of-mass momentum points along the
$Q_{x}$ or $Q_{y}$ axis \citep{Hu2025PRB,Hong2025}, mirroring the
behavior of the two-electron bound state.

\begin{figure}
\begin{centering}
\includegraphics[width=0.5\textwidth]{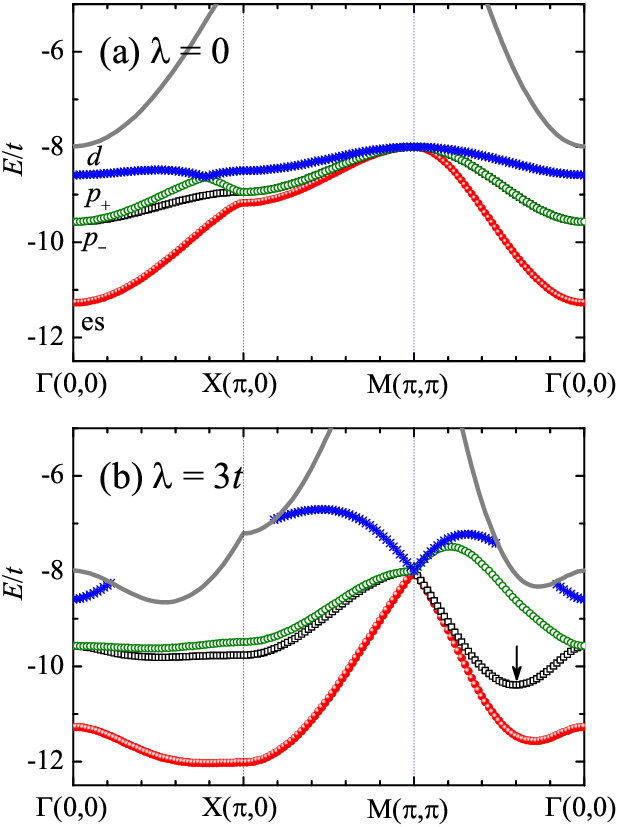}
\par\end{centering}
\caption{\label{fig3} The energy of the bound pairs as a function of the center-of-mass
momentum $\mathbf{Q}=(Q_{x},Q_{y})$ along the line cut $\Gamma$-$X$-$M$-$\Gamma$,
at the nearest-neighbor attractive interaction $V=-8t$, without (upper
panel) and with the $d_{xy}$-wave altermagnetic coupling $\lambda=3t$
(lower panel). Here, we take the on-site interaction $U=2t$. A strong
nearest-neighbor attraction gives rise to four bound states. In order
of increasing energy, these are represented by red filled circles,
black open squares, green open circles, and blue crosses. At the $\Gamma$
point, these states can be clearly classified as an extended $s$-wave
state, a twofold-degenerate $p$-wave state, and a $d$-wave state.
The arrow in the lower panel marks the minimum of the second-lowest
bound state along the $\Gamma$-$M$ direction.}
\end{figure}

\begin{figure*}
\begin{centering}
\includegraphics[width=0.75\textwidth]{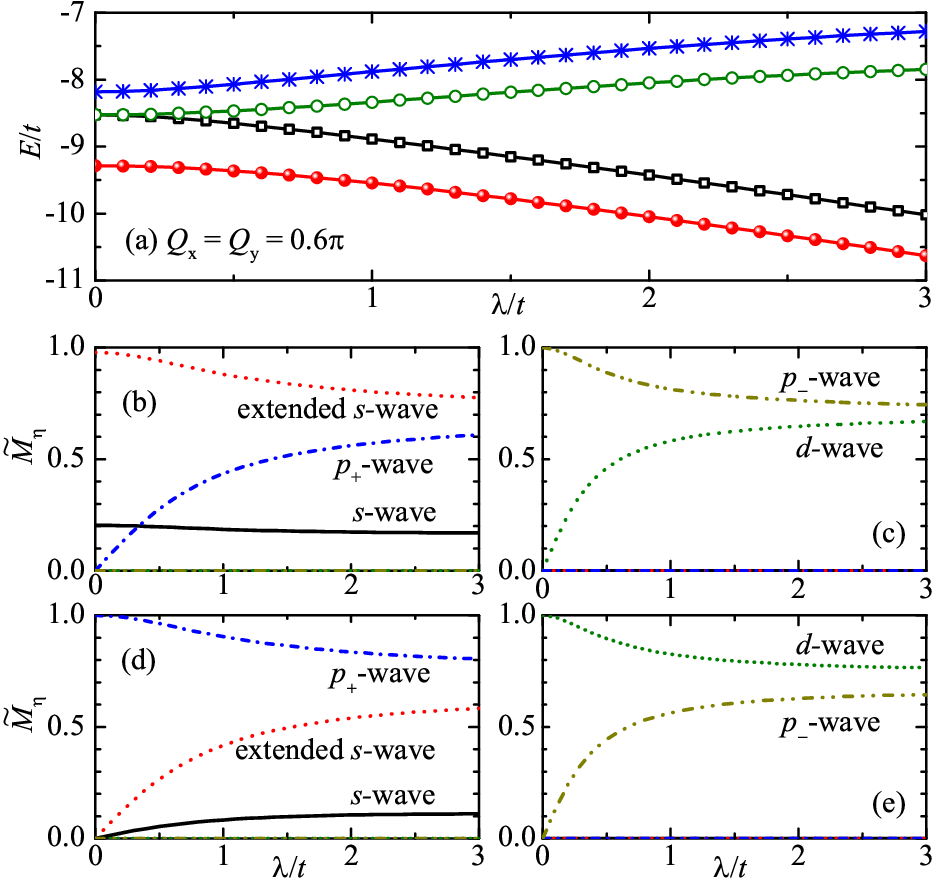}
\par\end{centering}
\caption{\label{fig4} (a) The energy of the bound pairs at $\mathbf{Q}=(Q_{x},Q_{y})=(0.6\pi,0.6\pi)$
as a function of the $d_{xy}$-wave altermagnetic coupling. The two
$p$-wave states split with increasing altermagnetic coupling. The
weight of the partial-wave components ($\tilde{M}_{\eta}$, as explicitly
indicated) of the four bound pairs: the ground-state pair (a), the
first-excited pair (b), the second-excited pair (c) and the third-excited
pair (d). For any nonzero altermagnetic coupling, we find a coexistence
of singlet partial-wave components ($s$-wave, extended $s$-wave
and $d$-wave) and and triplet partial-wave components ($p_{+}$-wave
and $p_{-}$-wave). Here, we take the on-site interaction $U=2t$
and the nearest-neighbor interaction $V=-8t$, as in Fig. \ref{fig3}.}
\end{figure*}

\subsection{The case with the nearest-neighbor attractive interaction $V<0$}

In this scenario, the nearest-neighbor attractive interaction can
support up to four distinct bound states \citep{Kornilovitch2004}.
As illustrated in Fig. \ref{fig3}(a), without altermagnetism the
four bound pairs at the $\Gamma$ point can be clearly identified
as extended $s$-wave, doubly degenerate $p$-wave, and $d$-wave
states \citep{Kornilovitch2004,Kornilovitch2024}. Moving away from
the $\Gamma$ point, the energies of these different bound states
may intersect at certain special momenta; however, their wave-functions
$\Phi_{\mathbf{k}}$ always remain either even or odd in $\mathbf{k}$
(not shown), which forces that the bound pairs are either spin-singlet
or spin-triplet. Remarkably, at the $M$ point all the four bound-pair
energies become degenerate. This degeneracy again arises from the
$\eta$-pairing mechanism discussed earlier: the two electrons occupy
nearest-neighbor sites and pair in four different configurations,
all with the same bound-state energy $E=V=-8t$.

The inclusion of an altermagnetic coupling $\lambda=3t$ leads to
several notable changes in the bound-pair dispersion. First, similar
to the case of on-site attraction, the lowest bound-pair energy with
domiant develops a global minimum at a finite center-of-mass momentum
along the $Q_{x}$ or $Q_{y}$ axis, following the lower edge of the
two-electron continuum. The second lowest bound-pair energy also acquires
a global minimum at nonzero momentum, but along the diagonal direction
(i.e., $Q_{x}=Q_{y}$ along the $\Gamma$-$M$ line, as highlighted
by the arrow in Fig. \ref{fig3}(b)). Therefore, in a many-electron
system, one would anticipate an FFLO superconducting state with a
finite center-of-mass momentum $\mathbf{Q}$. Second, because of the
pair-breaking effect of altermagnetism, we find that the fourth bound
state, marked by blue stars, ceases to exist over a range of momenta.
With increasing altermagnetic coupling strength, additional bound
states progressively disappear (not shown). Finally and most importantly,
except at the stationary $\Gamma$ and $M$ points with the highest
symmetry, the bound states are no longer degenerate, In particular,
the doubly degenerate $p$-wave states along the $\Gamma$-$M$ line
in the absence of altermagnetism are now split. As we discuss below,
this splitting arises from the mixed character of the bound states:
they can no longer be cleanly classified as spin-singlet or spin-triplet. 

\begin{figure*}
\begin{centering}
\includegraphics[width=0.25\textwidth]{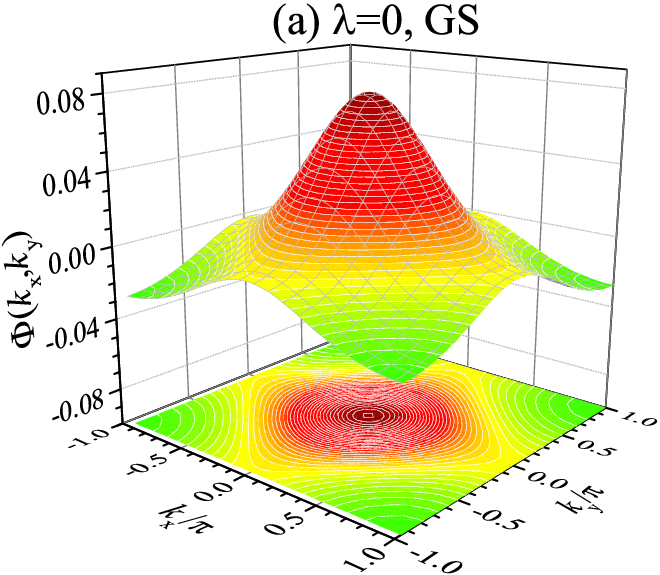}\includegraphics[width=0.25\textwidth]{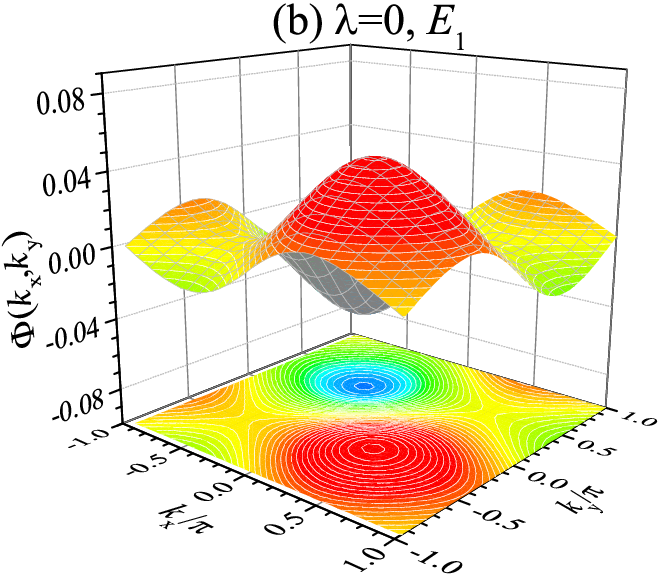}\includegraphics[width=0.25\textwidth]{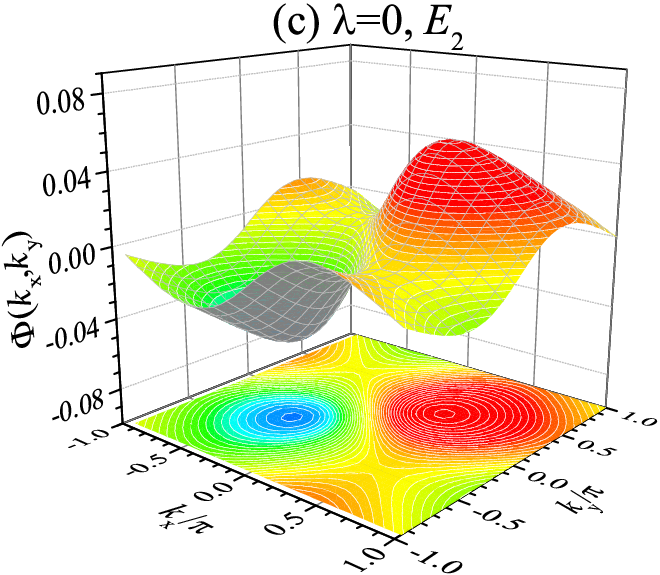}\includegraphics[width=0.25\textwidth]{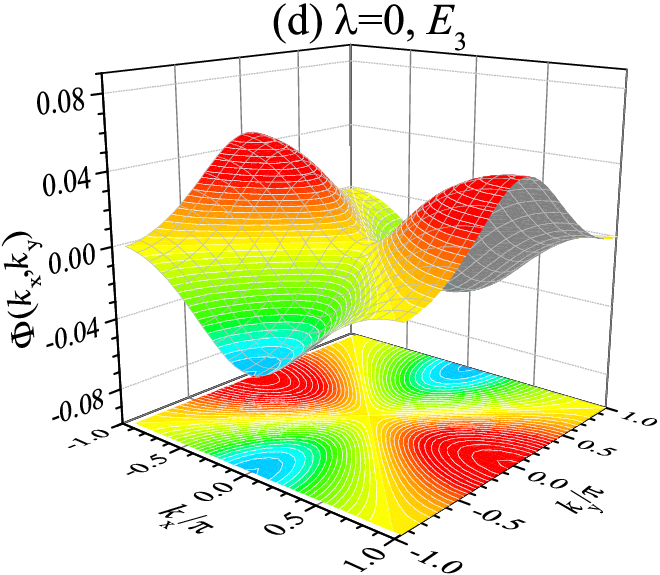}
\par\end{centering}
\begin{centering}
\includegraphics[width=0.25\textwidth]{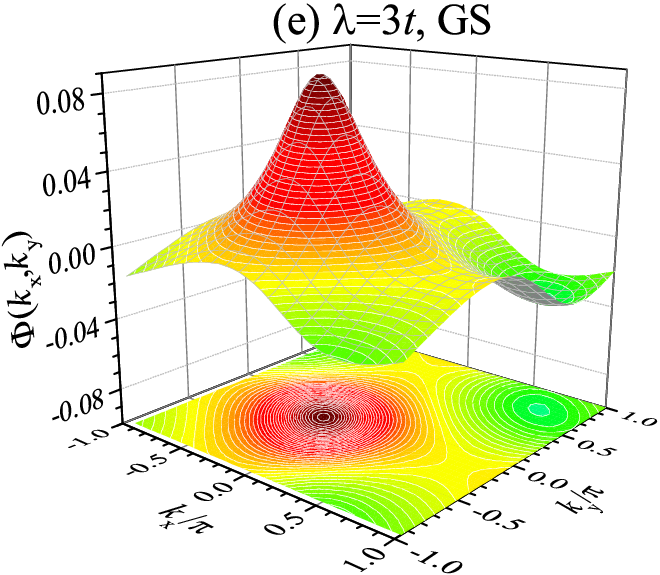}\includegraphics[width=0.25\textwidth]{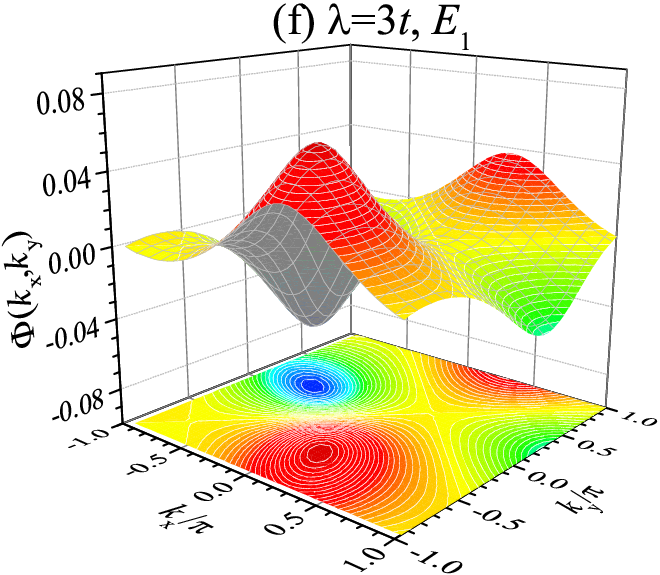}\includegraphics[width=0.25\textwidth]{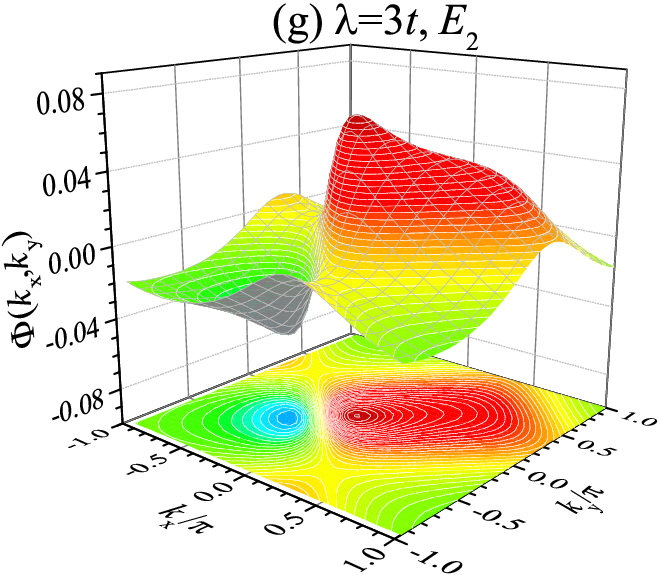}\includegraphics[width=0.25\textwidth]{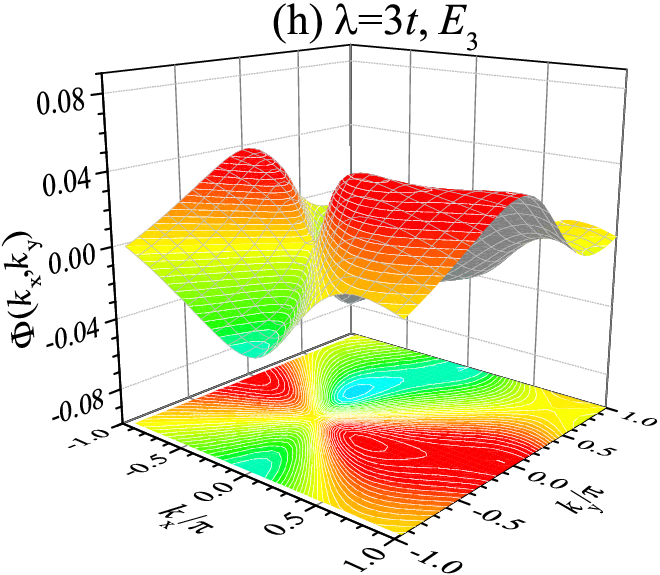}
\par\end{centering}
\caption{\label{fig5} The wave-functions $\Phi_{\mathbf{k}}$ of the four
bound pairs at $\mathbf{Q}=(Q_{x},Q_{y})=(0.6\pi,0.6\pi)$, without
(upper panel) and with the $d_{xy}$-wave altermagnetic coupling $\lambda=3t$
(lower panel), ordered by increasing energy. Introducing altermagnetic
coupling ($\lambda=3t$ ) significantly reshapes the momentum distribution
of each wave function and breaks rotational symmetry. Here, we take
the on-site interaction $U=2t$ and the nearest-neighbor interaction
$V=-8t$, as in Fig. \ref{fig3}.}
\end{figure*}

\subsection{spin-singlet and spin-triplet mixing}

In Fig. \ref{fig4}(a), we plot the energy splitting of the four bound
pairs as a function of the altermagnetic coupling $\lambda$ at a
center-of-mass momentum $\mathbf{Q}=(0.6\pi,0.6\pi)$. The corresponding
evolution of the different partial-wave components for each bound
state is presented in Figs. \ref{fig4}(b)-\ref{fig4}(e), ordered
by increasing bound-pair energy. Here, we choose the momentum $\mathbf{Q}$
along the diagonal direction, where the singlet-triplet mixing is
found to be most pronounced across the entire set of four bound-pair
states.

For the lowest bound state in Fig. \ref{fig4}(b), the dominant pairing
symmetry is aways the extended $s$-wave. As the altermagnetic coupling
increases, the $p_{+}$-wave component grows rapidly and becomes appreciable
by $\lambda=3t$. There is also a small residual $s$-wave contribution
induced by the on-site repulsion $U=2t$, depending only weakly on
the altermagnetic coupling. Therefore, the ground-state bound pair
clearly forms a coherent superposition of spin-singlet (extended $s$-wave)
and spin-triplet ($p_{+}$-wave) components. Analogous to bonding
and anti-bonding molecular states, this superposition implies the
existence of a complementary anti-bond state in which the roles of
the extended $s$-wave and $p_{+}$-wave components are reversed.
From Fig. \ref{fig4}(d), we see that this anti-bond state is exactly
given by the third lowest bound pair, where the state that was initially
$p_{+}$-wave dominated evolves, at sufficiently strong altermagnetic
coupling, into a nearly equal-weight superposition of $p_{+}$-wave
and extended $s$-wave. Similarly, Fig. \ref{fig4}(c) and Fig. \ref{fig4}(e)
show that the second and fourth bound pairs correspond to the bonding
and anti-bonding combinations of the $p_{-}$-wave and $d$-wave components,
respectively.

We have also examined the evolution of singlet-triplet mixing as the
altermagnetic coupling $\lambda$ increases for a center-of-mass momentum
directed along the $x$-axis, $\mathbf{Q}=(0.8\pi,0)$, where the
global ground-state bound pair is realized at $\lambda=3t$. In this
configuration, the relevant $p$-wave channel is described by the
form factor $f_{p_{y}}=\sqrt{2}\sin(k_{y})$, which corresponds to
an equal anti-bonding superposition of the $p+$ and $p-$ form factors.
For the lowest bound-pair state, the $p$-wave contribution grows
monotonically with increasing $\lambda$ and reaches its maximum weight
at $\lambda=3t$.

The mixed character of the bound pairs can also been clearly seen
in their pairing wave-function $\Phi_{\mathbf{k}}$, as reported in
Fig. \ref{fig5}. In the absence of altermagnetism (see the upper
panel), each momentum-space wave-function exhibits a well-defined
symmetry: extended $s$-wave, $p$-wave or $d$-wave symmetry, with
increasing bound-state energy. By contrast, introducing the altermagnetic
coupling $\lambda=3t$ substantially alters the momentum distribution
of each wave-function, breaking rotational symmetry. For instance,
in the ground-state bound pair illustrated in Fig. \ref{fig5}(e),
the wave-function maximum shifts aways from the origin and a pronounced
minimum emerges near the corner of first Brillouin zone at the $M$
point, reflecting the admixture of the $p$-wave component. 

It is worth noting that, at $\mathbf{Q}=(Q_{x},Q_{y})=(0.6\pi,0.6\pi)$
there is a mirror symmetry $m_{d}$ along the diagonal $k_{x}=k_{y}$.
Under this symmetry, the $s$-wave, extended $s$-wave, and $p_{+}$-wave
channels are even, while the $p_{-}$-wave and $d$-wave channels
are odd. As $E_{2p}(\mathbf{k},\mathbf{Q})$ remains unchanged under
the exchange $k_{x}\leftrightarrow k_{y}$ when $Q_{x}=Q_{y}$ (see
Eq. (\ref{eq:E2pLAM})), only pairing channels with the same mirror
parity can hybridize. Consequently, the $s$-wave, extended $s$-wave,
and $p_{+}$-wave components can mix with one another, as shown in
Fig. \ref{fig4}(b) and Fig. \ref{fig4}(d), and similarly the $p_{-}$-wave
and $d$-wave channels mix among themselves.

The observed singlet-triplet mixing is expected to have significant
consequences for the finite-momentum FFLO superconducting state that
may emerge in a many-electron system. In that context, the many-particle
pairing predominantly occurs near the Fermi surface. At low electron
densities (or small lattice filling factor), one can anticipate that
the order parameter will involve a mixture of extended $s$-wave and
$p_{+}$-wave pairings. At higher electron densities, the presence
of of Fermi surface and the Pauli exclusion principle suppress pairing
at small relative momenta, making a mixed $d$-wave and $p_{-}$-wave
pairing more likely to characterize the order parameter.

\section{Conclusions and outlooks}

In summary, we have obtained an exact solution for two-electron bound
pairs in an altermagnetic metal on a square lattice, considering both
on-site and nearest-neighbor attractive interactions. The altermagnetic
coupling strongly alters the threshold of the two-electron scattering
continuum, producing a global minimum at finite center-of-mass momentum
when the coupling is sufficiently strong. The bound-pair energy closely
follows this modified continuum threshold, so that the ground state
of the bound pairs can occur at nonzero center-of-mass momentum regardless
of the form of the interaction. Our findings suggest that the finite-momentum
bound pair could represent a two-electron manifestation of the pairing
tendency associated with the proposed altermagnetism-driven FFLO superconducting
state.

By analyzing the relative wave function of bound pairs in momentum
space, we have further found that for a nearest-neighbor attraction
capable of supporting multiple partial-wave interactions, the finite-momentum
bound states exhibit a mixed pairing symmetry. Consequently, in a
many-electron system one should expect an FFLO order parameter with
a mixed singlet-triplet character. Notably, a recent mean-field study
has indeed reported such mixed singlet-triplet finite-momentum pairing
with both $d$-wave and $p$-wave components \citep{Jasiewicz2026},
in the presence of altermagnetism and nearest-neighbor attraction.
A comprehensive investigation of the pairing instability in many-electron
systems could be pursued in future work using the well-established
Thouless criterion under the ladder approximation within the many-body
$T$-matrix approach \citep{Thouless1960,Liu2006,Hu2019}.

Another promising direction for future research is to elucidate the
relationship between the symmetry of bound states and the symmetry
of the underlying lattice. In many-electron systems, this issue may
be closely connected to the pairing symmetry of superconducting states,
which is fundamentally constrained by lattice symmetries. For the
standard zero-momentum pairing, the classification of admissible pairing
symmetries can be systematically carried out using group-theoretical
methods \citep{Sigrist1991}. Extending such analyses to finite-momentum
pairing states, however, is considerably more challenging and requires
a more rigorous and comprehensive theoretical framework.
\begin{acknowledgments}
This research was supported by the Australian Research Council's (ARC)
Discovery Program, Grants Nos. DP240101590 (H.H.), FT230100229 (J.W.)
and DP240100248 (X.-J.L.).
\end{acknowledgments}

\appendix

\section{The spectrum of all the two-electron energy levels}

\begin{figure}
\begin{centering}
\includegraphics[width=0.5\textwidth]{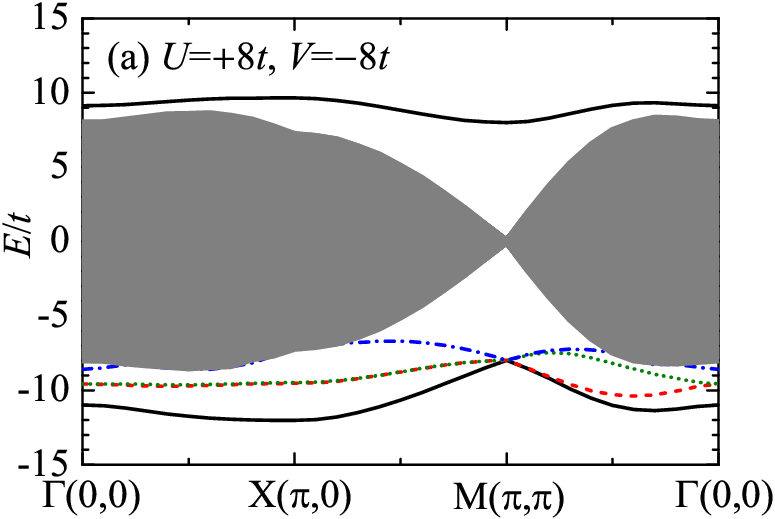}
\par\end{centering}
\begin{centering}
\includegraphics[width=0.5\textwidth]{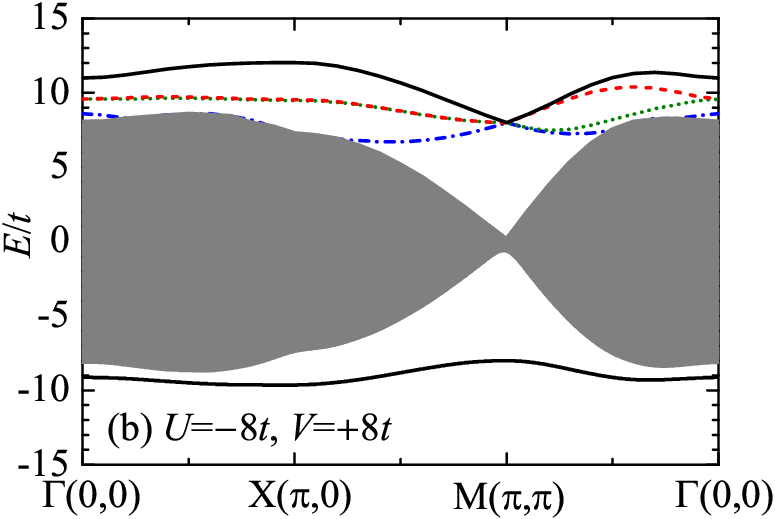}
\par\end{centering}
\caption{\label{fig6} The energy-level diagrams of two interacting fermions
with unlike spin. In the presence of the large on-site interaction
$U$ and nearest-neighbor interaction $V$, there are in general five
bound pairs, either at the bottom (for attractive interactions) or
at the top (for repulsive interactions) of the two-fermion continuum,
which are illustrated in grey. Here, we take a $d_{xy}$-wave altermagnetic
coupling $\lambda=3t$.}
\end{figure}

In Fig. \ref{fig6}, we show the complete two-electron spectrum along
the $\Gamma$-$X$-$M$-$\Gamma$ path, including both on-site and
nearest-neighbor interactions. For sufficiently strong interactions,
five bound states always appear, due to the existence of the five
distinct partial-wave channels. The sign of the interaction does not
affect this: attractive interactions produce bound pairs at the bottom
of the spectrum, while repulsive interactions produce bound pairs
at the top. At equal magnitudes of interaction strength, the attractive
and repulsive bound states are symmetrically positioned about $E=0$,
reflecting the particle--hole symmetry of the single-particle dispersion
$\varepsilon_{\mathbf{k}}$.

\section{The case of $d_{x^{2}-y^{2}}$-wave altermagnetism}

\begin{figure}
\begin{centering}
\includegraphics[width=0.5\textwidth]{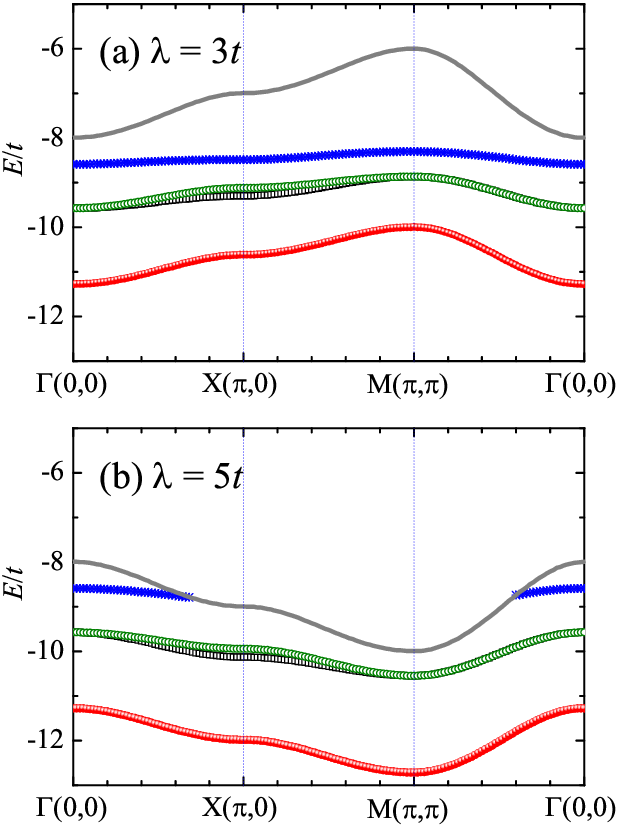}
\par\end{centering}
\caption{\label{fig7} The energy of the bound pairs as a function of the center-of-mass
momentum $\mathbf{Q}=(Q_{x},Q_{y})$ along the line cut $\Gamma$-$X$-$M$-$\Gamma$,
at the nearest-neighbor attractive interaction $V=-8t$, with the
$d_{x^{2}-y^{2}}$-wave altermagnetic coupling $\lambda=3t$ (a) and
$\lambda=5t$ (b). Here, we take the on-site interaction $U=2t$. }
\end{figure}

Here, we briefly summarize the behavior of bound pairs in the presence
of $d_{x^{2}-y^{2}}$-wave altermagnetism, where $J_{\mathbf{k}}=\lambda(\cos k_{x}-\cos k_{y})/2$.
The numerical calculations are performed by using the $p$-wave form
factors, $f_{\textrm{\ensuremath{p_{x}}}}(\mathbf{k})=\sqrt{2}\sin k_{x}$
and $f_{p_{y}}(\mathbf{k})=\sqrt{2}\sin k_{y}$. For $d_{x^{2}-y^{2}}$-wave
altermagnetism, the altermagnetic coupling $\lambda=4t$ turns out
to be a critical coupling constant, at which the single-particle dispersions
for spin-up and spin-down electrons becomes $\varepsilon_{\mathbf{k}\uparrow}=-4t\cos(k_{y})$
and $\varepsilon_{\mathbf{k}\downarrow}=-4t\cos(k_{x})$, resepctively.
It is then straighforward to see that the lower edge of the two-electron
continuum becomes flat, i.e., $E_{2p}^{(0)}(\mathbf{Q})=-4t$, regardless
of the center-of-mass momentum $\mathbf{Q}$.

In Fig. \ref{fig7}, we report the bound-pair energies for two $d_{x^{2}-y^{2}}$-wave
altermagnetic coupling strengths, $\lambda=3t$ (upper panel) and
$\lambda=5t$ (lower panel). Once $\lambda>4t$, a global minimum
develops at the $M$ point in the lower edge of the two-electron continuum,
supporting finite-momentum bound pairs at the same $M$ point. In
a many-eletron system, we anticipate the possible formation of a finite-momentum
FFLO-type order parameter. However, because of many-particle correlation,
the center-of-mass momentum is not necessarily pinned to the $M$
point $\mathbf{Q}=(\pi,\pi)$.

\end{document}